\documentclass[%
 reprint,
superscriptaddress,
 amsmath,amssymb,
 aps,
prb,
]{revtex4-1}
\usepackage[T1]{fontenc}
\usepackage{graphicx}
\graphicspath{{./images/}}
\usepackage{dcolumn}
\usepackage{bm}
\usepackage{siunitx}
\newcommand{\au}{a.u.}
\usepackage{xcolor}

\begin{document}


\title{Coulomb correlation in noncollinear antiferromagnetic $\alpha$-Mn}

\author{Aki Pulkkinen}
\email{aki.pulkkinen@lut.fi}
\affiliation{School of Engineering Science, LUT University, 53850 Lappeenranta, Finland}%

\author{Bernardo Barbiellini}%
\affiliation{School of Engineering Science, LUT University, 53850 Lappeenranta, Finland}
\affiliation{Department of Physics, Northeastern University, Massachusetts 02115 Boston, USA}%

\author{Johannes Nokelainen}
\affiliation{School of Engineering Science, LUT University, 53850 Lappeenranta, Finland}

\author{Vladimir Sokolovskiy}
\affiliation{Faculty of Physics, Chelyabinsk State University, 454001 Chelyabinsk, Russia}%
\affiliation{National University of Science and Technology ,``MISiS'', 119049 Moscow, Russia}

\author{Danil Baigutlin}
\affiliation{Faculty of Physics, Chelyabinsk State University, 454001 Chelyabinsk, Russia}%
\affiliation{School of Engineering Science, LUT University, 53850 Lappeenranta, Finland}

\author{Olga Miroshkina}
\affiliation{Faculty of Physics, Chelyabinsk State University, 454001 Chelyabinsk, Russia}%

\author{Mikhail Zagrebin}
\affiliation{Faculty of Physics, Chelyabinsk State University, 454001 Chelyabinsk, Russia}%
\affiliation{National Research South Ural State University, 454080 Chelyabinsk, Russia}

\author{Vasiliy Buchelnikov}
\affiliation{Faculty of Physics, Chelyabinsk State University, 454001 Chelyabinsk, Russia}%
\affiliation{National University of Science and Technology ,``MISiS'', 119049 Moscow, Russia}

\author{Christopher Lane}
\author{Robert S. Markiewicz}
\author{Arun Bansil}
\affiliation{Department of Physics, Northeastern University, Massachusetts 02115 Boston, USA}%

\author{Jianwei Sun}
\affiliation{Department of Physics and Engineering Physics, Tulane University, Louisiana 70118 New Orleans, USA}%

\author{Katariina Pussi}
\affiliation{School of Engineering Science, LUT University, 53850 Lappeenranta, Finland}

\author{Erkki L\"ahderanta}
\affiliation{School of Engineering Science, LUT University, 53850 Lappeenranta, Finland}

\date{\today}

\begin{abstract}
We discuss the interplay between magnetic and structural degrees of freedom in elemental Mn. The equilibrium volume is shown to be sensitive to magnetic interactions between the Mn atoms. While the standard generalized-gradient-approximation underestimates the equilibrium volume, a more accurate treatment of the effects of electronic localization and magnetism is found to solve this longstanding problem. Our calculations also reveal the presence of a magnetic phase in strained $\alpha$-Mn that has been reported previously in experiments. This new phase of strained $\alpha$-Mn exhibits a noncollinear spin structure with large magnetic moments.
\end{abstract}

\maketitle


\section{Introduction}
Manganese is one of the most complex metallic elements \cite{hobbs2001,hobbs2003,hafner2003,enteshami2017,ehteshami2018,lee2014,kolesnikov2016} that assumes many different stable crystal phases. On cooling the liquid, the sequence of crystal phases~\cite{bigdeli2015,kublerbook,proult1995} obtained includes body-centered cubic (BCC) $\delta$-Mn, face-centered cubic (FCC) $\gamma$-Mn, $\beta$-Mn, and $\alpha$-Mn as illustrated in Fig.~1. $\alpha$-Mn has 58 atoms per unit cell with space group $T^3_d$ (No. 217) \cite{bennett1987} and it may be looked upon as an intermetallic involving Mn atoms in different electronic and magnetic configurations \cite{bradley1927} on four crystallographic sublattices (I, II, III and IV). Neutron diffraction experiments \cite{lawson1994} have shown that sublattices III and IV further split into two types (IIIa, IIIb, IVa and IVb) when the antiferromagnetic ordering is taken into account. Large, almost collinear magnetic moments reside on sites I and II, while substantially smaller and strongly canted moments are on sites III and IV \cite{lawson1994}.

\begin{figure}
\centering
\includegraphics[width=.6\linewidth]{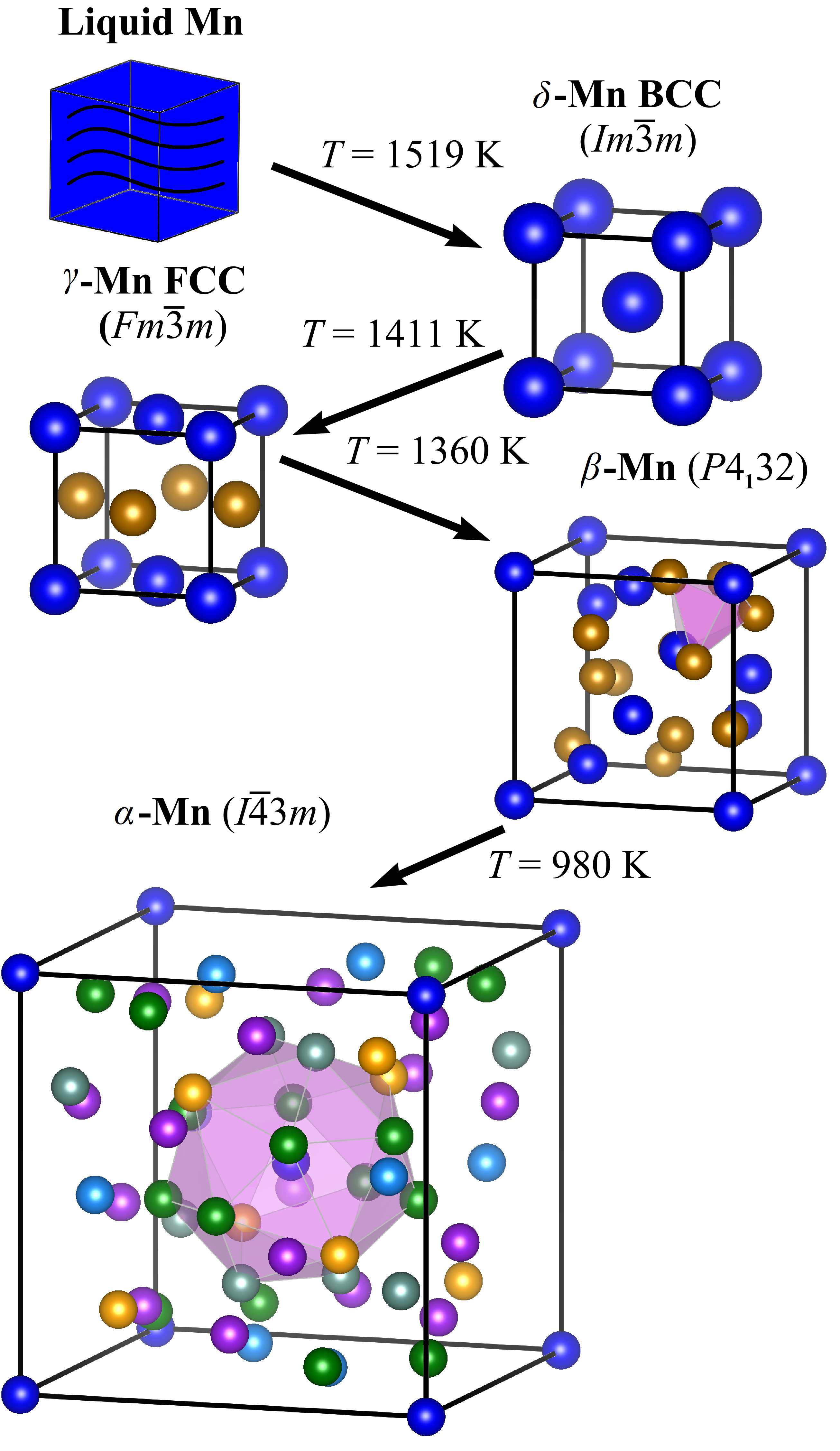}
\caption{\label{fig:mn_phases}(Color online) A schematic illustrating the evolution of various crystal phases of Mn from liquid Mn. Colors are used to differentiate between different Mn atoms in the unit cell. The polyhedron in the $\alpha$-phase structure is the 17-atom cluster described by Proult \& Donnadieu~\cite{proult1995}.}
\end{figure}

Density functional theory (DFT) results by Hobbs {\em et al.}~\cite{hobbs2001,hobbs2003} using either the local-spin-density approximation (LSDA) or the generalized-gradient approximation (GGA)  are not in agreement with experiments because the tendency of LSDA and GGA to overbind produces a collinear spin structure at the theoretical equilibrium volume. However, a noncollinear spin structure develops when the lattice is expanded beyond the experimental volume~\cite{hobbs2003}. 
A semiempirical tight-binding method using Hubbard-like correlation effects~\cite{sasaki1983} has predicted a noncollinear magnetic structure for $\alpha$-Mn at the experimental volume in qualitative agreement with experiment, but a more recent tight-binding study failed to converge to the noncollinear solution~\cite{suss1993}.

In order to address the deficiencies of the LSDA and GGA, exchange-correlation corrections must be improved. One approach is to introduce an ad hoc Hubbard parameter $U$~\cite{huang2018,liechtenstein1995,dudarev1998,cococcioni2005,ricca2019}, which attempts to correct for self-interaction errors on localized $3d$ orbitals of Mn by replacing LSDA and GGA potentials with orbital-dependent terms. An important test in this connection is the Mn$_2$ dimer~\cite{huang2018} which is discussed in the Supplemental Material (SM)~\cite{SM} (see also references~\cite{vosko1980,barth1972,moruzzibook,sprkkr} therein). Note that a DFT+$U$ approach requires both Hubbard-repulsion $U$ and Hund-exchange integral $J$ as ad hoc parameters \cite{yang2001}. Here the strongly-constrained-and-appropriately-normed (SCAN) functional~\cite{sun2015}, which is a semi-local functional that satisfies seventeen exact constraints, provides a systematic improvement over the GGA. It should be noted that SCAN leads to overestimated magnetic moments in itinerant ferromagnetic transition metals such as iron~\cite{isaacs2018,ekholm2018,fu2018}. However, there are many studies of antiferromagnetic materials such as the cuprates~\cite{lane2018,furness2018,zhang2018}, spinel LiMn$_2$O$_4$ cathode material~\cite{hafiz2019} and $3d$ perovskite oxides~\cite{varignon2019} where SCAN yields a good estimate of magnetic moments. This has also been shown to be the case in some Mn-rich Heusler alloys~\cite{buchelnikov2019,barbiellini2019}, where the $3d$ magnetic electrons are quite localized on the Mn atoms.

In this paper, we show that SCAN significantly improves the description of the ground-state electronic and magnetic structure of $\alpha$-Mn with respect to the LSDA and GGA by correctly accounting for conflicting trends for maximizing the magnetic spin moment and the bond strength. In this way, SCAN successfully captures the complex charge and noncollinear magnetic ordering that occurs in $\alpha$-Mn at low temperatures.

\section{Computational methods}
The present DFT calculations were performed with the plane-wave method implemented in the Vienna Ab Initio Simulation Package (VASP) \cite{kresse1993,kresse1996a,kresse1996b} with the projector augmented wave (PAW) method \cite{kresse1999}. The GGA exchange-correlation functional is based on the Perdew-Burke-Ernzerhof (PBE) formulation \cite{perdew1996} while the meta-GGA follows the SCAN implementation \cite{sun2015}. 
Structural relaxations were performed with an energy cutoff of $\geq \SI{550}{\electronvolt}$ and a k-point spacing of $< \SI{0.02}{\per\angstrom}$. The Methfessel-Paxton smearing method \cite{methfessel1989}  was used with a width of \SI{0.2}{eV} in geometry optimization runs, and the tetrahedron smearing method with Bl\"ochl corrections \cite{blochl1994} was used in self-consistency cycles as well as for generating the electronic density of states (DOS). Total energies were converged to \SI{e-6}{eV}. In geometry optimizations, forces on all atoms were converged to \SI{e-2}{\electronvolt\per\angstrom}. Spin polarization effects and the variational freedom for noncollinear spin arrangements were included for the $\alpha$-Mn structure. Note that the inclusion of noncollinearity in calculations significantly increases the computational cost as the electron density becomes a $2\times 2$ matrix~\cite{hobbs2000,zeleny2009}.

\section{Results}
We first examine $\gamma$-Mn with four Mn atoms per unit cell (see SM~\cite{SM} for details). Asada and Terakura \cite{asada1993} have shown that the LSDA underestimates the lattice constant and fails to predict the antiferromagnetic ground state. Our GGA and SCAN total energy calculations for the non-magnetic, ferromagnetic, and antiferromagnetic AF1 and AF2 phases confirm that the ground state of $\gamma$-Mn is AF1, where the sign of the moment alternates between the planes stacked along the $[001]$ direction. These phases are described by Kubler in Ref.~\cite{kublerbook}. GGA gives the Wigner-Seitz radius corresponding to the equilibrium volume \footnote{The Wigner-Seitz radius $R_{\rm ws}$ is related to the volume per Mn atom $V$ by: $V = 4\pi/3 R_{\rm ws}^3$.} of $R_{\mathrm{ws}}=\SI{2.635}{\au}$, while SCAN gives $R_{\mathrm{ws}}=\SI{2.732}{\au}$, which is in better agreement with the experimental value of $R_{\mathrm{ws}}=\SI{2.752}{\au}$~\cite{endoh1971}. The structure is found to be tetragonally distorted with $c/a = \num{0.95}$ for GGA, while SCAN produces $c/a = \num{0.98}$. The calculated magnetic moments increase with the equilibrium volume, thus GGA yields \SI{1.74}{\mu_B} whereas SCAN gives a higher value of \SI{3.10}{\mu_B}.  

\begin{figure}
\centering
\includegraphics[width=\linewidth]{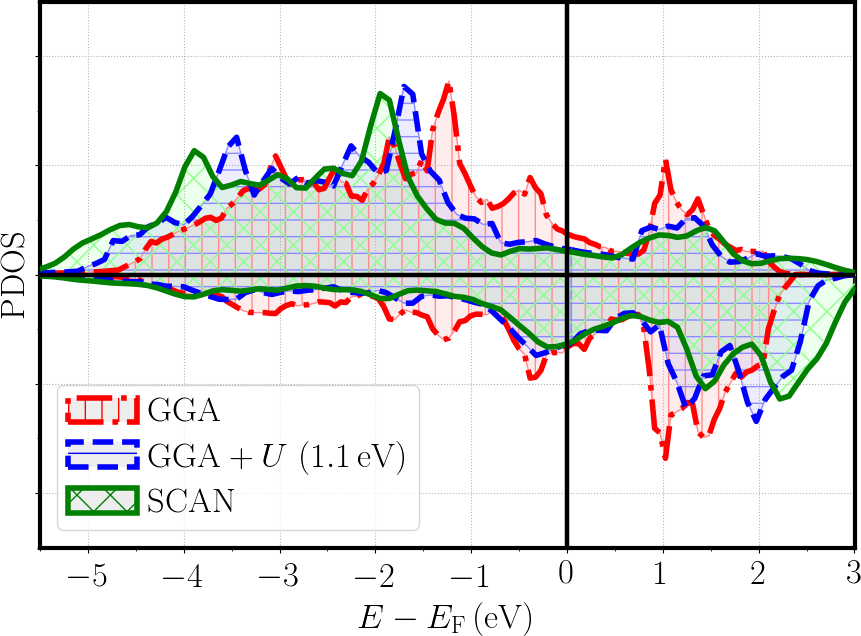}
\caption{\label{fig:fcc_dos}(Color online) Partial $l$-decomposed DOS (PDOS) for $d$-electrons in $\gamma$-Mn at the experimental $R_{\text{ws}} = 2.753$ a.u.\@ for GGA, GGA+$U$ (with $U = \SI{1.1}{eV}$) and SCAN.}
\end{figure}

In order to gauge the strength of corrections beyond the GGA captured by SCAN, we compare SCAN, GGA and GGA+$U$ results, and extract an effective $U$ value which reproduces the experimental equilibrium volume. In this way, we find that for $U = \SI{1.1}{eV}$, the equilibrium Wigner-Seitz radius is $R_{\mathrm{ws}}=\SI{2.722}{\au}$ with a Mn magnetic moment of \SI{2.69}{\mu_B} and $c/a=\num{0.98}$. These results are consistent with those reported previously by Podloucky \& Redinger~\cite{podloucky2018} and Di Marco \emph{et al.}~\cite{dimarco2009a,dimarco2009b}. Figure~\ref{fig:fcc_dos} highlights the effect of $U$ on the Mn partial DOS (PDOS) at the experimental volume. The GGA PDOS is seen to differ significantly from SCAN, but for $U=\SI{1.1}{eV}$ the two Mn PDOSs becomes closer. These results are also consistent with the observation of Hubbard bands in $\gamma$-Mn with angle-resolved photoemission spectroscopy~\cite{biermann2004}.

\begin{figure}
\includegraphics[width=\linewidth]{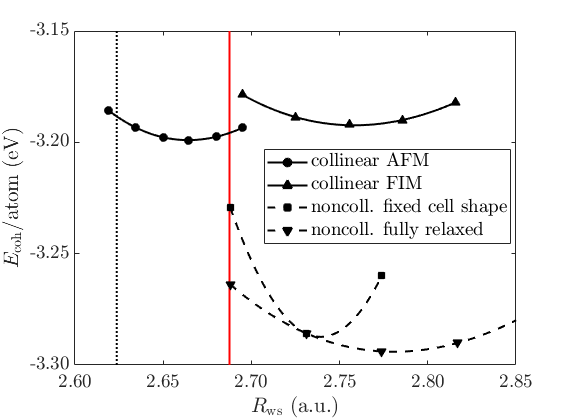}
\caption{\label{fig:mn_collinear}(Color online): SCAN-based cohesive energies for various magnetic structures of $\alpha$-Mn discussed in the text. The dotted vertical line corresponds to the GGA volume calculated with PBE~\cite{perdew1996} while the solid vertical line marks the experimental $R_{\text{ws}} = 2.688$ a.u.\@ of noncollinear $\alpha$-Mn~\cite{lawson1994}. $E_{\text{coh}}$ of $\gamma$-Mn in SCAN is $\SI{-3.27}{\electronvolt}$, see also the SM~\cite{SM}.}
\end{figure}

In $\alpha$-Mn, sublattices I, II and IV are occupied by close-packed polyhedra although this is not the case for sublattice III \cite{frank1958,bennett1987}. This close packing results in shrinking of the Wigner-Seitz radius to the experimental value \cite{lawson1994} of $R_{\mathrm{ws}}=\SI{2.688}{\au}$ Figure~\ref{fig:mn_collinear} shows the cohesive energy $E_{\text{coh}}$ as a function of $R_{\text{ws}}$ for a collinear solution. Even with this constraint SCAN significantly corrects the GGA volume. Moreover, the calculated bulk modulus of \SI{131}{GPa} is in good agreement with the experimental value of \SI{151}{GPa} \cite{fujihisa1995}. The computed magnetic moment distribution follows trends consistent with the experimental results given in Table II of Ref.~\cite{hobbs2003}. However, exact comparisons with theory are difficult because the moments extracted from the experimental data depend sensitively on the choice of the form factors used~\cite{kasper1956,oberteuffer1968,kunitomi1969, yamada1970, yamagata1972,lawson1994}. Interestingly, our computations predict the existence of an additional collinear (metastable) solution at a higher volume with a bulk modulus of \SI{68}{GPa}. The magnetic distribution for this solution becomes weakly ferrimagnetic with an average moment of \SI{0.11}{\mu_B} per Mn atom and the corresponding charge distribution involves Mn atoms in six different electronic configurations with charge differences reaching 0.4\,$e/\text{atom}$. When spin-orbit coupling (SOC) is included in the calculations, a noncollinear magnetic structure develops to reduce frustration. We have performed two different types of noncollinear calculations in this connection. The first involved fully relaxed atomic positions with a fixed cell-shape\footnote{In the fixed cell-shape calculations, the atomic positions were allowed to relax but the overall cell shape ($c/a$ ratio) was kept fixed at the experimental value ($c/a = 0.99955$). When both the cell shape and atom positions are allowed to relax, the c/a-ratio becomes 1.002.}, while the second achieved full structural relaxation.  The corresponding cohesive energies $E_{\text{coh}}$ as a function of $R_{\text{ws}}$ are shown in Fig.~\ref{fig:mn_collinear}. 

\begin{figure}
\includegraphics[width=.9\linewidth]{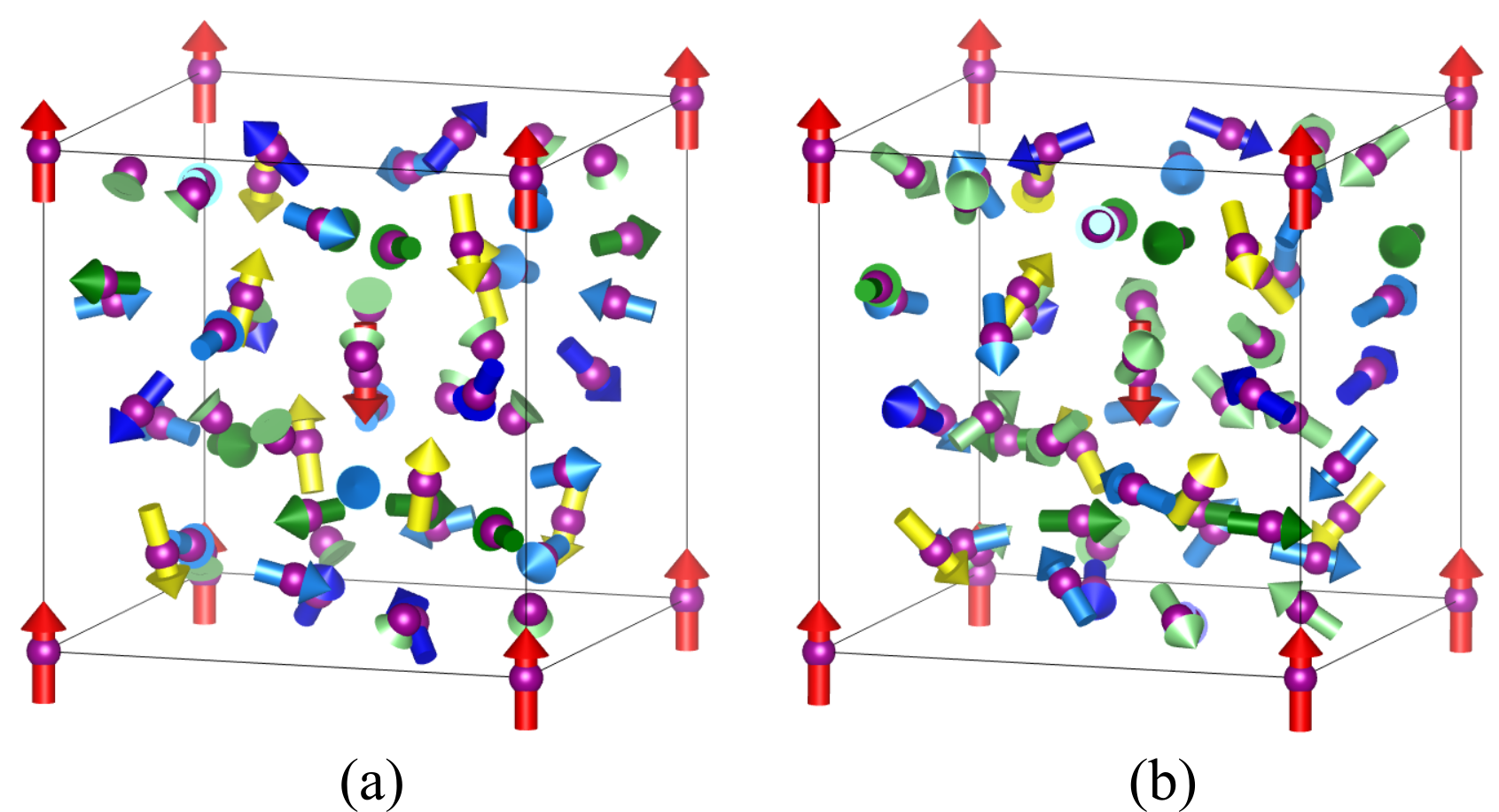}
\caption{\label{fig:mn_noncollinear}(Color online): SCAN-based noncollinear magnetic structure of various $\alpha$-Mn sublattices: I (red), II (yellow), IIIa (blue), IIIb (light blue), IVa (green), and IVb (light green). Only the atomic positions are relaxed in (a), while both the atomic positions and cell shape are relaxed in (b).
}
\end{figure}

In the GGA the structure remains collinear at the experimental volume \cite{hobbs2003} but in SCAN the moments rotate out of their collinear orientations. [The SCAN-generated magnetic structure is noncollinear for both the calculations shown in Fig.~\ref{fig:mn_noncollinear}.] In contrast to GGA, SCAN thus predicts noncollinear magnetic ordering at the experimental volume. In fact, as we noted above, we obtain two distinct magnetic structures, both with large collinear magnetic moments on Mn I sites, while the moments on Mn II sites are slightly smaller and canted away from the collinear direction. For the first solution, which is based on the fully-relaxed structure, we obtain $R_{\text{ws}} = 2.781$ a.u., and this solution might correspond to the strained $\alpha$-phase reported experimentally by Dedkov \emph{et al.}~\cite{dedkov2010}. The second solution involves computations with a fixed cell shape. Here the determination of the equilibrium volume is more delicate since several degenerate solutions with different spin structures can coexist as is the case in YBa$_2$Cu$_3$O$_7$~\cite{zhang2018} cuprate high-Tc superconductor. We note, however, that since SCAN tends to favor solutions with large magnetic moments, the stabilization of the strained $\alpha$-phase we have found might be due to the exaggerated corrections in SCAN~\cite{buchelnikov2019,mejia2019}. Noncollinear magnetism in manganese nanostructures has been reported also within the GGA~\cite{zeleny2009}.

The present noncollinear implementation of VASP~\cite{hobbs2000,zeleny2009} neglects noncollinear correlation effects beyond the LSDA~\cite{eich2013,pittalis2017}, which could possibly be the reason that SCAN yields solutions with too large equilibrium volume in $\alpha$-Mn\footnote{Interestingly, it is possible to stabilize a noncollinear solution for $\gamma$-Mn within SCAN. The collinear SCAN calculation yields $R_{\rm ws} = 2.732$ a.u. while the noncollinear calculation gives $R_{\rm ws} = 2.755$ a.u., which is in better agreement with the experimental value of $R_{\rm ws} = 2.752$ a.u. [The PBE value is $R_{\rm ws} = 2.635$ a.u.] Notably, the collinear SCAN result is only 3 meV/atom more stable than our noncollinear solution. Increase in the lattice constant for the noncollinear case is only 0.8\%. The lattice expansion with noncollinearity in $\gamma$-Mn is thus not large.}, see Fig.~\ref{fig:mn_collinear}. At the GGA level, a noncollinear spin structure stabilizes when the lattice is expanded beyond the experimental volume. In fact, in order to converge towards a noncollinear solution, Hobbs \emph{et al.}~\cite{hobbs2003} had to start their calculations with a strongly expanded initial volume ($R_{\rm ws} \approx 2.96$ a.u.), which however was not the final converged volume. The preceding results suggest that SCAN is a step in the right direction for stabilizing noncollinear solutions. 

Overmagnetization in $3d$ transition metals~\cite{isaacs2018,ekholm2018,fu2018} has been found in recent SCAN calculations. This issue has been addressed with a deorbitalized potential~\cite{mejia2019}. A similar problem is present in DFT+$U$, where the magnetic moment changes considerably if one adopts either the fully-localized limit or the mean-field approximation~\cite{stojic2008}. Here also a deorbitalized potential for states at the Fermi level has been suggested as a possible cure~\cite{barbiellini2005}.

\section{Summary and Conclusions}
Our results are relevant for smart materials such as the shape-memory and magnetocaloric Mn-rich Heusler alloys \cite{entel2018,buchelnikov2011,planes2009} because elemental Mn and the Mn-rich Heusler alloys present phase diagrams with common features. For example, BCC $\delta$-Mn and FCC $\gamma$-Mn can be viewed as austenite and martensite phases of Heusler alloys, respectively.  Although we have shown previously \cite{buchelnikov2019} that SCAN corrections beyond the GGA are small for Mn-poor compounds, our results here indicate that we can expect substantial differences between the GGA and SCAN in Mn-rich compounds such as $\mathrm{Ni}_2 \mathrm{Mn}_{1+x} (\mathrm{Ga}, \mathrm{Sn})_{1-x}$. SCAN corrections should work particularly well for short Mn-Mn distances where antiferromagnetic coupling tends to suppress itinerant ferromagnetism\cite{cardias2017}. In fact, the presence of spin- and charge-density wave like orderings in $\alpha$-Mn could help rationalize the complex phase diagrams of Heusler alloys and the associated phase instabilities driven by Fermi-surface nestings \cite{opeil2008,kimura2010,weber2015}. Since SCAN tends to promote complex solutions~\cite{zhang2018}, future investigations of Mn-rich materials should consider large simulation cells to capture modulated phases, which could be more stable than the simple martensitic~\cite{himmetoglu2012} phase. 

Our study provides a robust self-consistent scheme to correct the overbinding in elemental Mn in LSDA and GGA. The SCAN corrections for the equilibrium volume also yield noncollinear antiferromagnetism with complex charge and spin patterns in $\alpha$-Mn. These results demonstrate that the density-functional framework is capable of capturing the subtle correlation effects needed to predict technologically relevant Mn-rich materials for shape-memory, magnetocaloric and other applications.

\begin{acknowledgments}
It is a pleasure to acknowledge important discussions with John Perdew. The authors acknowledge CSC-IT Center for Science, Finland, for computational resources. The work of Chelyabinsk State University was supported by RSF-Russian Science Foundation project No. 17-72-20022 (calculations for $\gamma$- and $\delta$-Mn). The work at Northeastern University was supported by the US Department of Energy (DOE), Office of Science, Basic Energy Sciences grant number DE-FG02-07ER46352 (core research), and benefited from Northeastern University's Advanced Scientific Computation Center (ASCC), the NERSC supercomputing center through DOE grant number DE-AC02-05CH11231, and support (testing efficacy of advanced functionals) from the DOE EFRC: Center for Complex Materials from First Principles (CCM) under grant number DE-SC0012575. The work at Tulane University was supported by the startup funding from Tulane University. B.B. acknowledges support from the COST Action CA16218. 
\end{acknowledgments}

\bibliography{main.bib}
\end{document}